\documentclass[twocolumn,floatfix,prl,aps,showpacs,superscriptaddress]{revtex4}
\usepackage{epsfig,graphicx,tabularx}
\usepackage{subfigure}
\usepackage{amsmath,amsfonts,amssymb}
\usepackage{color,braket}

\begin{document}

\title{Spreading of correlations in a quenched repulsive and attractive one dimensional lattice system} 
\author{L. Barbiero} 
\affiliation{CNR-IOM DEMOCRITOS Simulation Center and SISSA, Via Bonomea 265, I-34136 Trieste, Italy}
\affiliation{Dipartimento di Fisica e Astronomia "Galileo Galilei", Universit\`a di Padova, 35131 Padova, Italy}
\author{L. Dell'Anna} 
\affiliation{Dipartimento di Fisica e Astronomia "Galileo Galilei", Universit\`a di Padova, 35131 Padova, Italy}

\begin{abstract}
We study the real time evolution of the correlation functions in 
a globally quenched interacting one dimensional lattice system by means of time adaptive density matrix renormalization group. 
We find a clear light-cone behavior quenching the repulsive interaction 
from the gapped density wave regime. The spreading velocity increases with the 
final values of the interaction and then saturates at a certain finite value. 
In the case of a Luttinger liquid phase as the initial state, for strong repulsive interaction quenches, a more complex dynamics occurs as a result of 
bound state formations. 
From the other side in the attractive regime, depending on where connected 
correlation functions are measured, one can observe a delay in the starting 
time evolution and a coexistence of ballistic and localized signals. 
\end{abstract}

\pacs{71.10.Fd, 03.75.Kk, 05.30.-d,}

\maketitle




\section{Introduction}
A clear understanding of out-of-equilibrium isolated quantum systems is one 
of the most challenging 
task in quantum physics \cite{Polkovnikov2011}. 
In this context a lot of effort has been devoted towards the comprehension of thermalization processes \cite{rigol,deutsch}, the role of conserved quantities \cite{rigol2,lucci}, entanglement dynamics \cite{coser, alba,daley} and many body-localization \cite{huse,vosk}.
A further key point is represented by the typical light-cone shape exhibited by one- and two-point correlation functions once a sudden quench is applied \cite{dechiara, lauchli, manmana, carleo,bonnes}. 
It has been showed \cite{calabrese} that in critical theories the maximum 
velocity of the spreading of correlations is given by twice the group velocity defined in the final gapless system. 
Actually the existence of a maximal velocity \cite{lieb,sims,kliesch}, known 
as the Lieb-Robinson bound, has been shown to exist theoretically in several 
locally interacting many-body systems. This is due to the short range interactions which 
may reduce the propagation of information making its spreading velocity finite.
Moreover light-cone propagation of correlations is expected when starting from a 
non-degenerate initial state that shows an exponentially cluster decomposition 
property \cite{bravyi,eisert1,kastner}, which is indeed generally valid for local 
Hamiltonians. The light cone propagation can be absent in the presence of long range interaction \cite{hauke,cevolani,eisert} or for some local spin models \cite{dellanna}. Nowadays, thanks to the impressive achievements in the field 
of ultracold systems \cite{Bloch2008} the aforementioned results have been tested by means of cold bosons \cite{cheneau,langen} and trapped ions \cite{jurcevic,richerme}.\\
Motivated by this intensive work activity in this paper we provide a time-dependent density matrix 
renormalization group (t-DMRG) \cite{feiguin} analysis of the correlation 
spreading once a sudden global quench is applied in system of spinless fermions. More precisely, in the first section we consider the case when the 
interaction is repulsive and we show how much the Bethe-Ansatz approach is 
able to properly capture the velocity of the excitation propagation.   
A further crucial point is given by the initial particles density 
distribution. 
Indeed, when we quench the interaction from a weak to a strong value, 
many-body bound states can 
give rise to a multi-signal propagation. 
In the attractive interaction regime, instead,
starting from a phase separated state, it possible to get 
two signals, a localized and a ballistic one. 
A crucial role is played by  
the position at which the correlation function is pinned. 
Indeed if one of the positions 
of the two-point correlation functions is taken 
well inside the occupied bulk, 
one has to wait a certain time 
before observing the quantum effect of the correlation spreading.
\begin{figure}[h]
\includegraphics[height=5cm,width=\columnwidth]{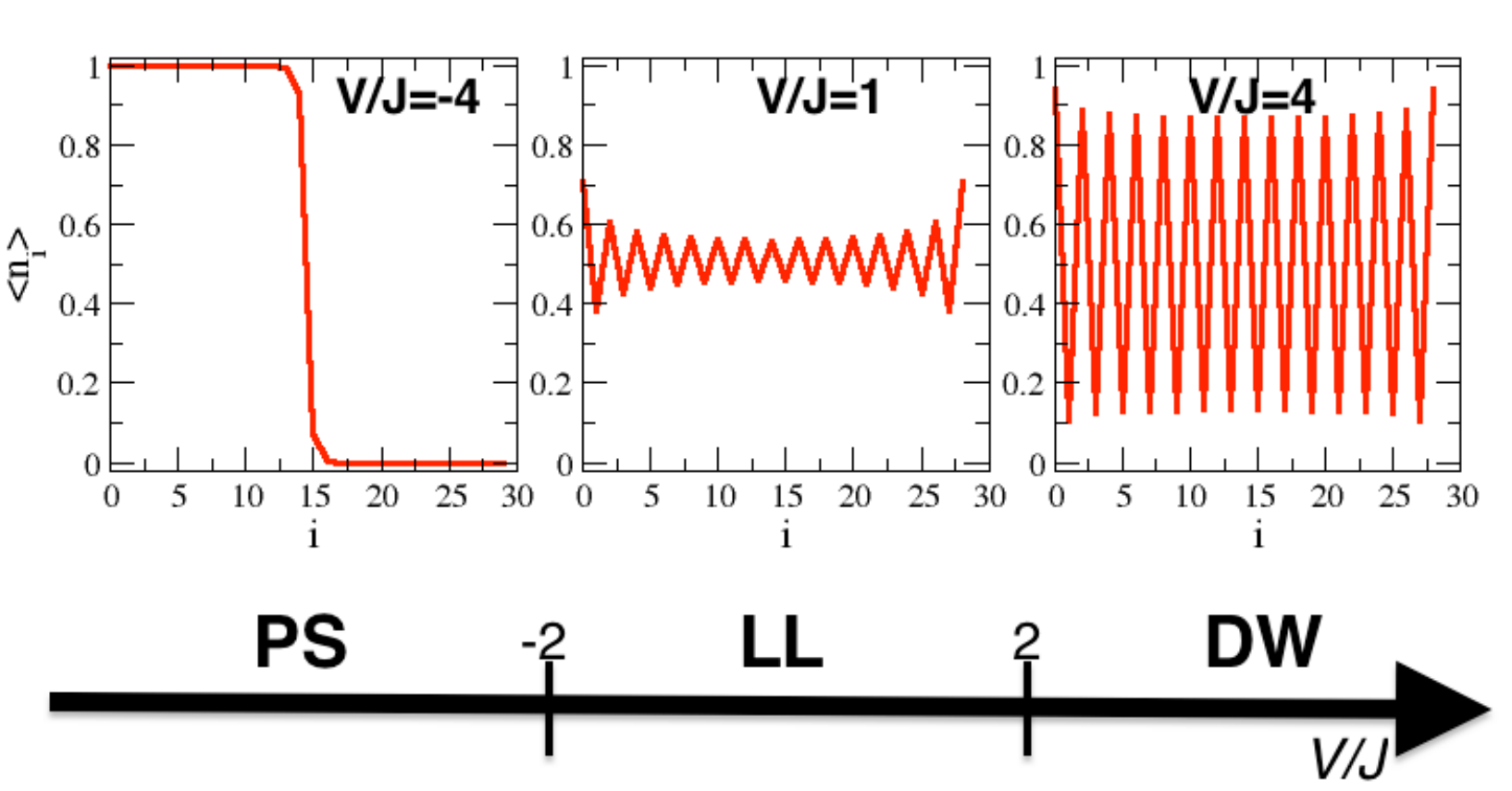}
\vspace*{-0.5cm}
\caption{(Color online) DMRG density profiles $\langle n_i\rangle$ for the three different regimes for system with $N=15$ bosons and $L=30$ sites. The degeneracy of the density wave (DW) is removed by considering $L=29$ sites. While the degeneracy of the phase separation state (PS) is broken by applying two chemical potentials $\mu/J$=0.001 with apposite sign at the lattice edges. The simulation are performed by keeping up to 512 DMRG states and 5 finite size sweeps.}
\label{fig1}
\end{figure}
\section{Model}
The system we consider is composed by $N$ spinless fermions loaded in $L$ sites of a one dimensional lattice at half filling $\bar{n}=N/L=1/2$
\begin{equation}
H=-J\sum_i\Big(c^\dagger_ic_{i+1}+c^\dagger_{i+1}c_{i}\Big)+V\sum_in_in_{i+1}
\label{ham}
\end{equation} 
where $J$ is the tunneling amplitude fixing our energy and time scales, $V$ 
is the nearest-neighbor (NN) interaction while $c^\dagger_i$ ($c_i$) is the 
creation (annihilation) operator of a particle in the $i$-site of the lattice. 
Notice that at half filling, i.e. $\bar{n}=1/2$, the Hamiltonian 
(\ref{ham}) turns out to be exactly mapped to the $XXZ$ spin-$1/2$ model by 
performing a Jordan-Wigner transformation. Remarkably the physics of 
the model Eq.~(\ref{ham}) can be basically studied in the experiments by using 
dipolar fermions/hard-core bosons \cite{Lahaye2009,baier}, bosonic mixture 
\cite{fukuhara} and photons \cite{gorshkov}. 
A further key feature of Eq.~(\ref{ham}) is given by its integrability 
\cite{baxter} which allows one to extract many fundamental properties. 
In particular it is well known that the phase diagram consists of three 
different phases: one of them being a gapless Luttinger liquid (LL) in the 
region $-2<V/J<2$ and two degenerate gapped regimes. The gap appears both for 
$V/J<-2$, giving rise to a phase separated (PS) state with empty and occupied 
sites totally demixed in two different regions and, and for $V/J>2$, where a density wave (DW) modulation is present. 
Notice that the two previous regimes are usually identified in spin language 
as ferro- and antiferro- magnetic regimes. 
\begin{figure}[h]
\includegraphics[scale=1.0]{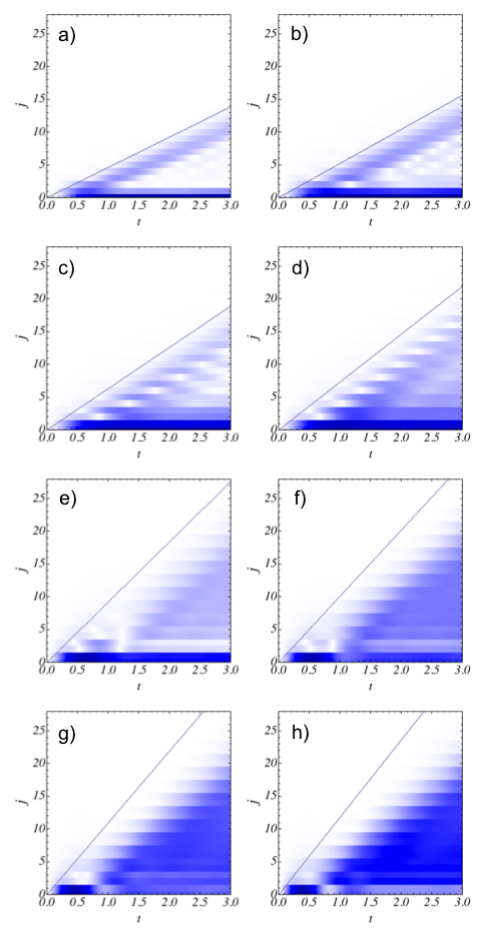}
\caption{(Color online) t-DMRG results of $\Delta C_{ij}=|C_{ij}(t)-C_{ij}(0)|$ for $V_i=4$ and a) $V_f=0.5$; b) $V_f=1$; c) $V_f=2$; d) $V_f=3$; e) $V_f=5$; f) $V_f=6$; g) $V_f=7$; h) $V_f=8$. The length of the chain is $L=29$ and $N=15$. Both $t$ and the $V$'s are in unit of $J$. 
The slops of the solid lines are the velocities $v_{BA}$, given by Eq.~(\ref{vba}). The static simulations for the ground state are performed by keeping up to 512 DMRG states and 5 finite size sweeps and the dynamics is obtained by using a time step $\delta=0.01$ and 250 DMRG states.}
\label{fig2}
\end{figure}
The relative density profiles $\langle n_i\rangle$, obtained via static DMRG 
calculations \cite{white} for the three different phase, are shown in 
Fig.~\ref{fig1}. Here we break the ground state degeneracy of the DW and PS 
phases by respectively considering an odd number of sites and by adding very 
small antiparallel chemical potentials at the lattice edges. 
As shown in \cite{bravyi,eisert1,kastner} degeneracy breaking is needed in 
order to observe physical time-spreading of the correlations. 
The weak density modulation appearing in LL regime is nothing but the well 
know phenomenon of the Friedel oscillations \cite{friedel}. 
Here the crucial feature, as it will be clear in the next part, is 
that the wave function of any single 
particle is sufficiently delocalized to allow to two particles to lie in NN sites. This is an obvious difference with respect to the DW regime where the 
strong repulsive $V$ makes energetically very costly NN occupancies. 
From the other side PS allows NN occupations due to obvious energetic reasons 
but the single particle wave function is localized. 
For this reason it is important to understand how a certain initial density 
distribution affects both the spreading of the correlation and its velocity 
once a sudden quench in the interaction $V$ is applied.
In all our simulations we obtain the ground state (GS) relative to a certain 
$V_i$ and we let the system evolve in time $t$ once the interaction is 
suddenly brought to a value $V_f$. We, then, monitor the relative spreading of 
the connected density-density correlation function
\begin{equation}
C_{ij}(t)=\langle n_i(t)n_j(t)\rangle-\langle n_i(t)\rangle\langle n_j(t)\rangle\,.
\label{eq1}
\end{equation}
\section{Repulsive Regime}
As a first step we investigate the case of a $V_i$ supporting a DW regime. 
In Fig. \ref{fig2} we plot $\Delta C_{ij}=|C_{ij}(t)-C_{ij}(0)|$ as a function 
of $t$ and $j$, at fixed $i=0$ (first site of the lattice), for $V_i=4$ and  
several values of $V_f$, both in the LL and DW phase. 
\begin{figure}[h]
\includegraphics[height=7cm,width=\columnwidth]{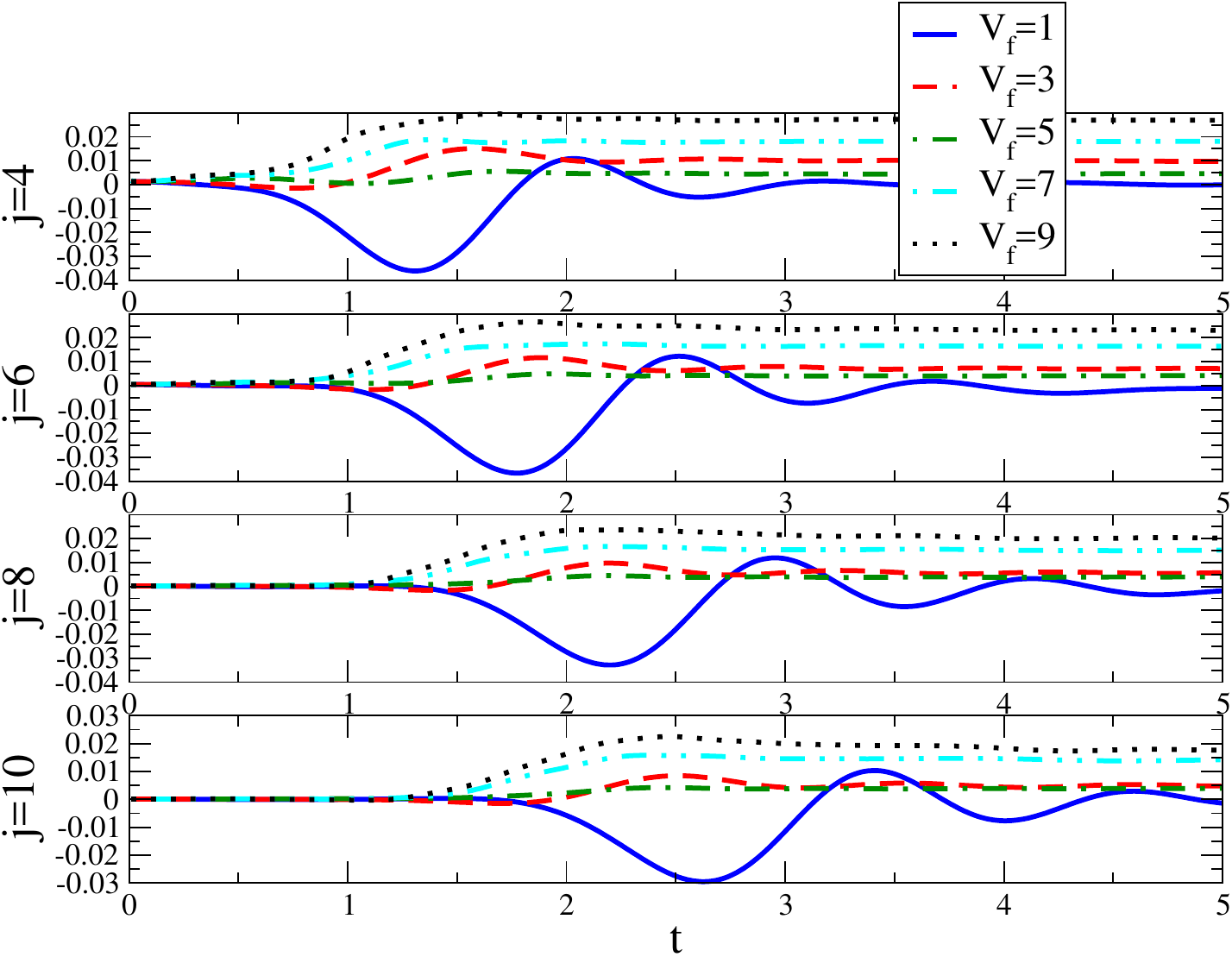}
\caption{(Color online)  t-DMRG results of $\Delta C_{ij}=C_{ij}(t)-C_{ij}(0)$ as a function of $t$ for $V_i=4$ and different values of $j$ and $V_f$. 
The length of the chain is $L=29$ and $N=15$. Both $t$ and the $V$'s are in unit of $J$. 
The static simulations for the ground state are performed by keeping up to 512 DMRG states and 5 finite size sweeps and the dynamics is obtained by using a time step $\delta=0.01$ and 250 DMRG states.}
\label{fig3}
\end{figure}
For $V_f$ within the gapless regime, the exact spectrum from Bethe-Ansatz 
approach is available \cite{cazalilla},  
which allows us to predict the spreading velocity $v$ in a gapless phase, which is $v\simeq v_{BA}$, with 
\begin{equation}
v_{BA}=2\pi J \frac{\sqrt{1-(V_f/2J)^2}}{\arccos(V_f/2J)}\,,
\label{vba}
\end{equation}
\begin{figure}[h]
\includegraphics[scale=1.0]{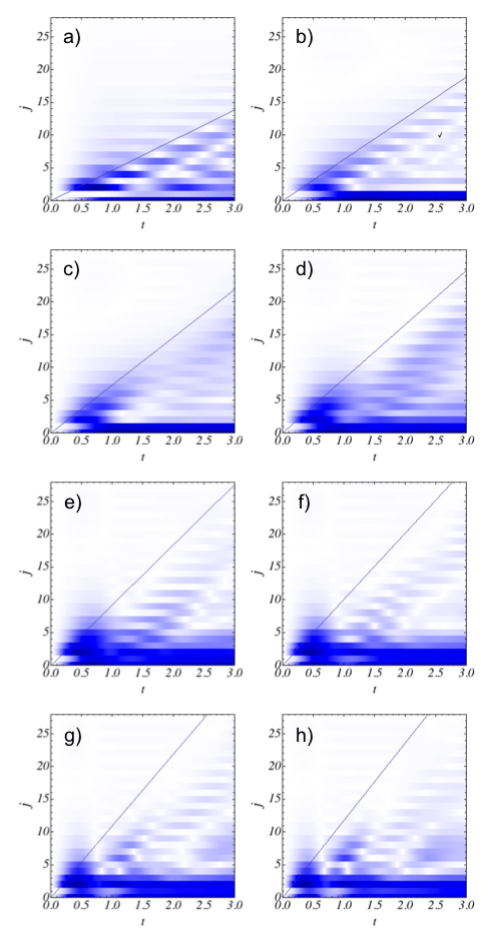}
\caption{(Color online) t-DMRG results of $\Delta C_{ij}=|C_{ij}(t)-C_{ij}(0)|$ for $V_i=1$ and a) $V_f=0.5$; b) $V_f=2$; c) $V_f=3$; d) $V_f=4$; e) $V_f=5$; 
f) $V_f=6$; g) $V_f=7$; h) $V_f=8$. 
The length of the chain is $L=30$ and $N=15$. Both $t$ and the $V$'s are in unit of $J$. The slops of the solid lines are the velocities $v_{BA}$,
given by Eq.~(\ref{vba}). 
The static simulations for the ground state are performed by keeping up to 512 DMRG states and 5 finite size sweeps and the dynamics is obtained by using a time step $\delta=0.01$ and 250 DMRG states.}
\label{fig4}
\end{figure}
twice the sound velocity. 
As clearly visible, our numerical results in Fig. \ref{fig2} are in a very 
good agreement with the analytical solution, for $V_f$ in the LL regime. 
Furthermore, as already shown in \cite{manmana}, for $V_f$ not too strong, but 
already able to capture the DW behavior, the Bethe-Ansatz velocity, extended to $V_f>2$, is still able to give a description of the spreading velocities. 
From the other side once $V_f$ exceeds a critical value, the velocity $v$ of the fastest signal seems to saturate to a constant value. In particular, as visible in Fig. \ref{fig3}, for $V_f\gtrsim 5$ the velocity saturates at the value $v\sim8$, in agreement with the results reported in \cite{manmana} for a different value of $V_i$. 
As we checked, the aforementioned feature remains valid also for different 
values of $V_i$, confirming the belief that the spreading velocity depends 
mainly on the final interaction strength governing the dynamics. Of course 
extracting the exact velocity $v$ is not possible within our approach so we 
can only infer on the correctness of the just mentioned results. However a key 
point to be noticed here is that only one clear and strong single signal is present in the dynamics.
The situation becomes rather different once interaction quenches are performed 
starting from a LL ground state. Several results showing quenches within this regions are presents, see \cite{collura} and references therein, but the case of strong $V_f$ has not been investigated. As previously mentioned in the gapless 
regime the density distribution is weakly modulated but the wave functions 
are rather delocalized thus allowing NN occupancies. As clearly visible in 
Fig.~\ref{fig4}, this aspect has huge consequences in excitations propagation. 
More precisely in Fig.~\ref{fig4} we start with a LL configuration obtained by 
getting the ground state for $V_i=1$ and we let the system evolve with 
different $V_f$. Clearly when $V_f$ is weak a clear single-signal propagation 
is visible, resembling the $V_i=4$ case, but once $V_f$ becomes stronger a 
multi-signal propagation appears. Moreover it is possible to notice that the 
bigger is $V_f$ the bigger is the number of signals contributing to the 
dynamics. This effect can be basically explained by starting with a 
two-particles description, looking at the two-body energy spectrum \cite{nguenang,ganahl}. Indeed, in 1D the energy spectrum of two interacting particles has 
a continuos set of scattering states and, for NN or on-site interaction, one 
bound state outside this region. 
If the interaction strength is strong enough the bound state is able to support the presence of localized solutions, like one-site bound pairs for on-site 
interaction \cite{zoller} or inter-site bound pairs for long-range interaction 
\cite{me}. To be more specific about our model Eq.~(\ref{eq1}), for $V_i=1$ 
two particles can lie in NN sites and once the interaction is suddenly brought 
to strong values, the systems is projected in a state with high energy. 
If such an energy corresponds to the one of a bound state, the system is not 
able to decay in the scattering region and remains trapped in an excited 
level. This is due to the fact that the system is isolated, i.e. the energy is conserved, and to 
the kinetic energy limitation induced by the lattice structure.
As a consequence the two particles form a NN inter-site bound pair which can tunnel with $J_{eff}\sim J^2/V$, namely much slower than the single particle 
tunneling processes. 
In Fig.~\ref{fig4} two signals are visible meaning that the energy we are 
providing to the system is not sufficient to form all the possible bound pairs
but only a fraction of particles are bounded while the rest 
behaves as single ones, with a tunneling amplitude $J$. 
The fact that for strong enough $V_f$ more than two signals 
contribute to the dynamics is because 
also three- and in general many-body bound states can be formed, see, for instance, Ref. \cite{valiente} for the case with on-site 
interaction and \cite{ganahl} for NN interactions. Of course, the bigger is 
the number of bounded particles the slower will be the velocity associated to its expansion.   
\section{Attractive Regime}
\begin{figure}[h]
\includegraphics[scale=1.0]{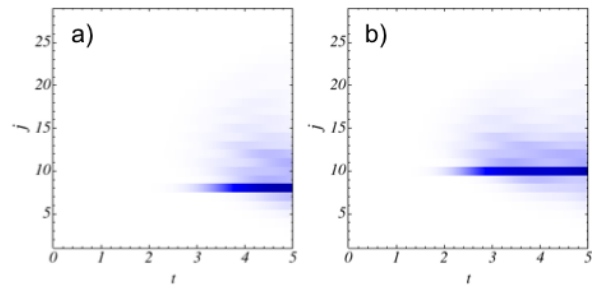}
\caption{(Color online) t-DMRG results of $\Delta C_{ij}=|C_{ij}(t)-C_{ij}(0)|$ for $V_i=-4$ and $V_i=-1$  with a) $i=8$ and  b) $i=10$. The chain length is $L=30$ and $N=15$. Both $t$ and the $V$'s are in unit of $J$. The static simulations for the ground state are performed by keeping up to 512 DMRG states and 5 finite size sweeps  and using two antiparallel chemical potentials $\mu/J$=0.001 in the lattice edges. In the dynamics the chemical potentials are removed and we use a time step $\delta=0.01$ and 250 DMRG states.}
\label{fig5}
\end{figure}
For $-2<V_i<0$ the density distribution is still almost constant. This feature combined with the fact that the two body energy spectrum for NN interacting particles 
is symmetric under the transformation $V\leftrightarrow -V$ makes intuitive to understand that, in analogy with LL-DW quench, the same 
multi-signal propagation is observed for $V_i$ in a LL regime and $V_f$ in the deep PS one. 
On the contrary, the dynamics driven by quenching from a PS to a LL regime is 
completely unexplored and, as we will see, gives rise to a very different scenario 
with respect to the repulsive case. Due to strongly attractive NN interaction, 
all the $N$ particles are compressed in $L/2$ lattice sites while the 
others are empty (see the first panel of Fig.~\ref{fig1}). 
Due to the Fermi statistic, the physical properties inside 
the occupied region are basically the same of a classical system. 
Indeed, if we perform a global interaction quench fixing the position $i$ of 
the connected density-density correlation function, Eq. (\ref{eq1}), well 
inside this region we do not observe any signal propagation for long time, 
namely the correlation function is fully classical until that a characteristic 
time is passed. 
Actually, for $V<-2$ the GS is highly degenerate, meaning that the real GS 
density distribution of a PS state is uniform \cite{note}. 
Consequently, if at $t=0$ we break the degeneracy by applying 
small local chemical potentials and we let the system evolve 
without them the system will try to restore the GS degeneracy by letting the 
high density region expand even if no interaction quenches are performed. 
Of course this case is different from the repulsive DW regime
where the two-fold GS degeneracy is broken by using an odd number of sites 
$L$. The crucial point is that the particles occupying the populated region need a certain time, 
which depends on the single particle 
position, to expand. More precisely, once the evolution begins, the particle 
located at the border between the occupied and empty regions can tunnel in 
one empty site thus letting another NN particle to tunnel, and so on. 
This kind of tunneling has a strong influence on the 
correlation spreading after an interaction quench. 
As a result what is very important for the dynamics is the position $i$ in the connected density-density correlation function Eq.~(\ref{eq1}). 
Indeed, if one pins $i=0$ one has to wait a long time before seeing any 
propagation signal due to the fact that all the other particles have to move 
first. 
\begin{figure}[h!]
\includegraphics[scale=1.0]{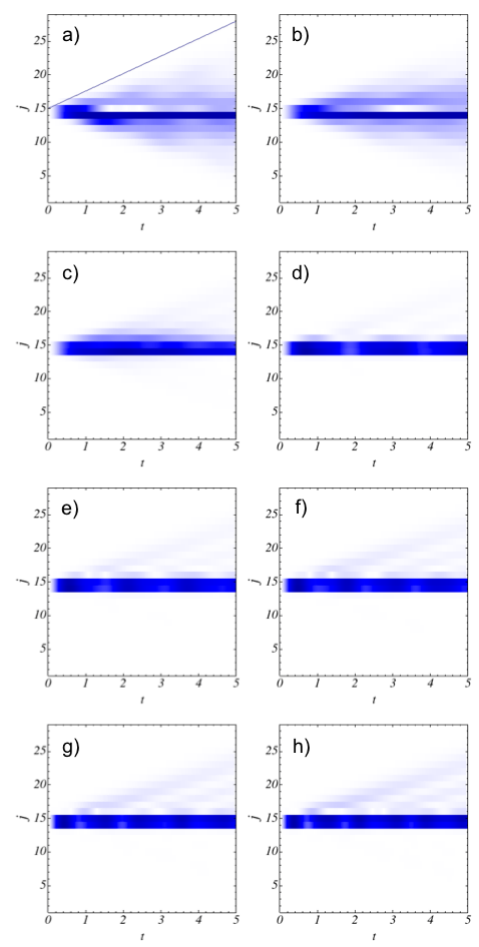}
\caption{(Color online) t-DMRG results of $\Delta C_{ij}=|C_{ij}(t)-C_{ij}(0)|$ for $V_i=-4$ and a) $V_f=-1$; b) $V_f=-2$; c) $V_f=-3$; d) $V_f=-5$; e) $V_f=-6$; f) $V_f=-7$; g) $V_f=-8$; h) $V_f=-9$. 
The length of the chain is $L=30$ and $N=15$. Both $t$ and the $V$'s are in unit of $J$. 
The slop of the solid lines in the first panel 
is the velocity $v_{BA}$, given by Eq.~(\ref{vba}).
The static simulations for the ground state are performed by keeping up to 512 DMRG states and 5 finite size sweeps  and using two antiparallel chemical potentials $\mu/J$=0.001 in the lattice edges. In the dynamics the chemical potentials are removed and we use a time step $\delta=0.01$ and 250 DMRG states.}
\label{fig6}
\end{figure}
For the same reason, if $i$ is pinned in a more central site the correlation propagation will appear sooner and clearly if $i$ is located exactly at the border between the two regions with different densities the propagation will start immediately. 
These features are visible in Fig.~\ref{fig5} where we fixed $i$ in different positions and the propagation starts at different times. 
As clearly shown in  Fig.~\ref{fig5}, 
for $i$ pinned in the $8$-th site the propagation begins 
later with respect to the case with $i=10$, namely in a site where the quantum 
nature of the system is restored earlier. 
This behavior proves that once a sort of classic nature is imposed in an 
initial state, one has to wait a characteristic time before pure quantum 
effects, i.e. connected correlation spreading, take place. 
Calling $i_0$ the position of the border of the separated regions, 
from our numerical results we can estimate this waiting time  
\begin{equation} 
t\sim (i_0-i)/3\,.
\end{equation}
In the examples shown in Fig.~\ref{fig5}, $i_0=15$ and $i=8$ (left panel) and $i=10$ (right panel), and the waiting times are respectively $t\approx 2.2$ and $t\approx 1.6$. 
Finally in Fig.~\ref{fig6} we plot the propagation for $i=i_0$, namely at the 
border of the occupied and empty lattice regions for the GS. 
In this case, as discussed before, the time evolution of the quantum signals 
starts immediately. 
As shown in Fig.~\ref{fig6}, starting from the PS phase, $V_i=-4$, 
and quenching in the LL we observe a quantum propagation spreading inwards and 
outwards the region which was previously highly occupied at $t=0$. The 
inner signal becomes weaker for $V_f<-2$ and almost disappears for $V_f<V_i$, 
namely quenching deep in the PS phase.  
In this regime, instead, there is still the outer propagation with a light-cone shape and another even stronger signal localized at the center of the system. 
This latter effect is probably due to the formation of a many-body bound state 
which corresponds to a cluster of $\sim N$ particles. Clearly due to the large 
number of particles this cluster has an expansion velocity which is very 
small and it explains why for our time scale the signal remains mainly 
localized at the center of the lattice. 
From the other side, in analogy with the repulsive regime, the Bethe-Ansatz prediction, Eq.~\ref{vba}, is able to roughly capture the light cone-like 
velocity propagation when the value of $V_f$ is deep in the LL phase, 
while it fails approaching the phase transition. The Bethe-Ansatz velocity $v_{BA}$, indeed, goes to zero for $V_f\rightarrow -2$, 
while this is not the case for the velocity $v$ of the correlation spreading, 
which is finite as shown in Fig.~\ref{fig6}, pannel b). 
Interestingly we get almost the same spreading velocity for any $V_f$ in 
the gapped phase and its value $v\sim 2.5$ is much smaller than that observed 
for the repulsive case. 
\section{Conclusions}
We studied the correlation spreading  in an interacting fermionic system 
after a sudden quench. We shown that depending on the initial condition, different excitation propagations can be observed. 
In particular, a quench from an initial density wave phase  
supports the presence of one clear light-cone signal. 
In this regime we found that the propagation velocity is, for a 
relatively large range of final interaction values, in good agreement with the 
Bethe-Ansatz predictions and then saturates to $v\sim 8$, for $V_f\gtrsim 5$.  
The situation is rather different if the initial state is prepared in the 
Luttinger liquid phase. In this regime delocalized wave functions can 
give rise to a multi-signal propagation for strong interacting quenches. 
These different velocities are associated to the formation of bound-states.  
Finally we studied the case when we start from a phase separated state. 
Here, depending on the points where the connected correlation 
functions are measured, the propagation signals can be delayed.  
Moreover, for strong quenches, together with a weak signal of a ballistic 
evolution, with a spreading velocity $v\sim 2.5$ independent on $V_f$, also a 
very slow and strong signal associated with many-body bound states, can be 
observed. 
As a last remark we stress that all our results can be proved in experiments involving either cold atoms or photons.      


{\it Acknowledgements.--}  We thank P. Calabrese for useful discussions. This work was supported by MIUR~(FIRB 2012, Grant No. RBFR12NLNA\_002). L.B. thanks the CNR-INO BEC Center in
Trento for CPU time.




\thebibliography{99}

\bibitem{Polkovnikov2011} A. Polkovnikov, K. Sengupta, A. Silva, and
  M. Vengalattore, Rev. Mod. Phys. {\bf 83}, 863 (2011).

\bibitem{rigol} M. Rigol, V. Dunjko, and M. Olshanii, Nature (London)
{\bf 452}, 854 (2008).

\bibitem{deutsch} J. M. Deutsch, Phys. Rev. A {\bf43}, 2046 (1991).

\bibitem{rigol2} M. Rigol, V. Dunjko, V. Yurovsky, and M. Olshanii, Phys.
Rev. Lett. {\bf 98}, 050405 (2007).

\bibitem{lucci} A. Iucci and M. A. Cazalilla, Phys. Rev. A {\bf 80}, 063619
(2009).

\bibitem{coser} A. Coser, E. Tonni, and P. Calabrese, J. Stat. Mech. (2014) P12017.

\bibitem{alba} V. Alba, and  P. Calabrese arXiv:1608.00614 .

\bibitem{daley} A. S. Buyskikh, M. Fagotti, J. Schachenmayer, F. Essler, A. J. Daley Phys. Rev. A {\bf 93}, 053620 (2016).

\bibitem{huse} R. Nandkishore, D. A. Huse, Annual Review of Condensed Matter Physics, Vol. 6: 15-38 (2015).

\bibitem{vosk} R. Vosk, D. A. Huse, E. Altman, Phys. Rev. X {\bf 5}, 031032 (2015).


\bibitem{dechiara} G. De Chiara, S. Montangero, P. Calabrese, and R. Fazio,
J. Stat. Mech.: Theory Exp. (2006) P03001.

\bibitem{lauchli} A. M. L\"auchli and C. Kollath, J. Stat. Mech.: Theory Exp.
(2008) P05018.

\bibitem{manmana} S. R. Manmana, S.Wessel, R.M. Noack, and A. Muramatsu,
Phys. Rev. B {\bf 79}, 155104 (2009).

\bibitem{carleo} G. Carleo, F. Becca, L. Sanchez-Palencia, S. Sorella, and
M. Fabrizio, Phys. Rev. A {\bf 89}, 031602 (2014).

\bibitem{bonnes} L. Bonnes, F. H. L. Essler, A. M. L\"auchli, Phys. Rev. Lett. {\bf 113}, 187203 (2014).


\bibitem{calabrese} P. Calabrese and J. Cardy, Phys. Rev. Lett. {\bf 96}, 136801 (2006).

\bibitem{lieb} E. H. Lieb and D.W. Robinson, Commun. Math. Phys. {\bf 28},
251 (1972).

\bibitem{sims} R. Sims and B. Nachtergaele, \textit{Lieb-Robinson Bounds in
Quantum Many-Body Physics}, edited by R. Sims and D.
Ueltschi, Entropy and the Quantum, Vol. 529 (American
Mathematical Society, Providence, RI, 2010).

\bibitem{kliesch} M. Kliesch, C. Gogolin, and J. Eisert, \textit{Lieb-Robinson
Bounds and the Simulation of Time Evolution of Local
Observables in Lattice Systems}, edited by L. D. Site and
Bach, Many-Electron Approaches in Physics, Chemistry
and Mathematics: A Multidisciplinary View (Springer, New
York, 2013).

\bibitem{bravyi} S. Bravyi, M. B. Hastings, and F. Verstraete, Phys. Rev. Lett. {\bf 97} 050401 (2006).

\bibitem{eisert1} J. Eisert, and T. J. Osborne, Phys. Rev. Lett. {\bf 97} 
150404 (2006).

\bibitem{kastner} M. Kastner, New, J. Phys. {\bf 17} 123024 (2015).

\bibitem{hauke} P. Hauke and L. Tagliacozzo, Phys. Rev. Lett. {\bf 111}, 207202
(2013).

\bibitem{eisert} J. Eisert, M. van den Worm, S. R. Manmana, and M.
Kastner, Phys. Rev. Lett. 111, 260401 (2013).

\bibitem{cevolani} L. Cevolani, G. Carleo, and L. Sanchez-Palencia, Phys. Rev. A {\bf 92}, 041603(R) (2015).

\bibitem{dellanna} L. Dell'Anna, O. Salberger, L. Barbiero, A. Trombettoni, and V. Korepin, arXiv:1604.08281; Phys. Rev. B, in press

\bibitem{Bloch2008} I. Bloch, J. Dalibard and W. Zwerger,
  Rev. Mod. Phys. {\bf 80}, 885 (2008).

\bibitem{cheneau} M. Cheneau, P. Barmettler, D. Poletti, H. Endres, P. Schau§,
T. Fukuhura, C. Gross, I. Bloch, C. Kollath, and S. Kuhr,
Nature (London) 481, 484 (2012).

\bibitem{langen} T. Langen, R. Geiger, M. Kuhnert, B. Rauer, and
J. Schmiedmayer, Nat. Phys. 9, 640 (2013).

\bibitem{jurcevic} P. Jurcevic, B. P. Lanyon, P. Hauke, C. Hempel, P. Zoller, R.
Blatt, and C. F. Roos, Nature (London) 511, 202 (2014).

\bibitem{richerme} P. Richerme, Z.-X. Gong, A. Lee, C. Senko, J. Smith, M.
Moss-Feig, S. Michalakis, A. V. Gorshkov, and C. Monroe,
Nature (London) 511, 198 (2014).

\bibitem{feiguin} S. R. White and A. E. Feiguin, Phys. Rev. Lett. {\bf 93}, 076401 (2004); A. E. Feiguin and S. R. White, Phys. Rev. B {\bf 72}, 020404(R) (2005).

\bibitem{Lahaye2009} See e.g. T. Lahaye, C. Menotti, L. Santos,
  M. Lewenstein, and T. Pfau, Rep. Prog. Phys. {\bf 72}, 126401
  (2009), and references therein.

\bibitem{baier} S. Baier, M. J. Mark, D. Petter, K. Aikawa, L. Chomaz, Zi Cai, M. Baranov, P. Zoller, and F. Ferlaino,  Science {\bf 352}, 201-205 (2016).

\bibitem{fukuhara} T. Fukuhara, P. Schau§, M. Endres, S. Hild, M. Cheneau, I. Bloch, and C. Gross, Nature 502, {\bf 76} (2013).

\bibitem{gorshkov} A. V. Gorshkov, J. Otterbach, E. Demler, M. Fleischhauer, M. D. Lukin, Phys. Rev. Lett. {\bf 105}, 060502 (2010).

\bibitem{baxter} R.J. Baxter, \textit{Exactly solved models in statistical mechanics}, London, Academic Press, 1982.

\bibitem{white} S. R. White, Phys. Rev. Lett. {\bf 69}, 2863 (1992).

\bibitem{friedel} J. Friedel, N. Cimento Suppl. {\bf 7}, 287 (1958).

\bibitem{cazalilla} M. A. Cazalilla, R. Citro, T. Giamarchi, E. Orignac, M. Rigol, Rev. Mod. Phys. {\bf 83}, 1405-1466 (2011).

\bibitem{collura} M. Collura, P. Calabrese, and F. H. L. Essler, Phys. Rev. B {\bf 92}, 125131 (2015).

\bibitem{nguenang}J.-P. Nguenang and S. Flach, Phys. Rev. A {\bf 80}, 015601 (2009).

\bibitem{ganahl} M. Ganahl, E. Rabel, F. H. L. Essler, and H. G. Evertz, Phys. Rev. Lett. {\bf 108}, 077206 (2012).

\bibitem{zoller} K. Winkler, G. Thalhammer, F. Lang, R. Grimm, J. Hecker Denschlag, A. J. Daley, A. Kantian, H. P. Buechler, P. Zoller, Nature {\bf 441}, 853 (2006).

\bibitem{me} L. Barbiero, C. Menotti, A. Recati, and L. Santos, Phys. Rev. B 92, 180406(R) (2015).

\bibitem{valiente} M. Valiente, D. Petrosyan, and A. Saenz,  Phys. Rev. A {\bf 81}, 011601(R) (2010).


\bibitem{note} The density would be exactly constant in presence of translational symmetry, i. e. in a ring geometry. In a box like geometry the density is almost constant in the center of the lattice showing oscillations at the edges.
\end{document}